\documentclass{article}
\usepackage{Style/arxiv}
\usepackage[utf8]{inputenc} 
\usepackage[T1]{fontenc}    
\usepackage{hyperref}       
\usepackage{url}            
\usepackage{booktabs}       
\usepackage{amsfonts}       
\usepackage{nicefrac}       
\usepackage{microtype}      
\usepackage{lipsum}		
\usepackage{graphicx}
\usepackage{doi}
\usepackage{hyperref}
\usepackage{bm}
\AtBeginDocument{%
  }

\usepackage{algpseudocode}
\usepackage{algorithm}
\usepackage{tikz-cd}
\usepackage{amsmath}
\usepackage{amsthm}
\usepackage{multirow}

\usepackage{xr}
\usepackage{float}
\usepackage{wrapfig}
    \usepackage{caption}
    \usepackage{subcaption}
\usepackage{listings}

\title{Decoupling Appearance and Geometry in Diffusion World Models}

\author{
Rajit Rajpal\\
University of Edinburgh \\
\texttt{s2592586@ed.ac.uk}
}
\renewcommand{\shorttitle}{Decoupling Appearance and Geometry in Diffusion World Modelss}

\hypersetup{
pdftitle={Decoupling Appearance and Geometry in Diffusion World Models},
pdfauthor={Rajit Rajpal},
}

\begin{document}
\maketitle
\begin{abstract}
We study the problem of learning diffusion-based world models that jointly capture scene geometry and visual appearance in complex first-person environments. Existing approaches typically train a single action-conditioned model that predicts future RGB frames, but they either ignore geometry or require action supervision for all modalities, limiting scalability to large unlabeled video datasets. In this work, we introduce a two-stage diffusion world model that separates geometric and appearance prediction while maintaining joint temporal coherence. Our key idea is a staggered RGB–depth formulation: depth at time $t+1$ is predicted first using an action-conditioned diffusion transformer, and the RGB frame at the same timestep is then generated conditioned on this predicted depth using a second diffusion model without action inputs. We show that this decoupled design faithfully approximates a hypothetical joint RGBD diffusion model trained with diffusion forcing, enabling the second stage to be trained or fine-tuned on additional RGB–depth data without requiring action labels. Experiments on the CS:GO Dust~2 dataset demonstrate that our approach improves both depth accuracy and RGB fidelity compared to fully joint models and RGB-only baselines, while significantly enhancing scalability to augmented or unlabeled datasets. Our results highlight the effectiveness of geometry-guided diffusion for world modeling and open a path toward large-scale action-free video model training. Code is available on \href{https://github.com/rajit906/Qubit_Oasis}{Github}.
\end{abstract}


\section{Introduction}

Modeling complex 3D environments from high-dimensional sensory observations is a central challenge in computer vision and embodied AI. Recent advances in diffusion models have dramatically improved our ability to synthesize realistic images, videos, and 3D structures. At the same time, world models have emerged as a powerful paradigm for learning predictive representations of environments, enabling agents to forecast future observations, reason about dynamics, and act effectively in interactive settings. However, building world models that are both temporally coherent and geometrically grounded remains an open problem. A key difficulty lies in jointly capturing visual appearance and scene geometry. Structural guidance (primarily from depth maps) has been shown to be very beneficial in video world modeling \cite{aether}, \cite{tesseract}, \cite{zahorodnii2025improvingworldmodelsusing}\cite{yan2025} \cite{chen2025deepverse4dautoregressivevideo}. While RGB observations provide rich photorealistic detail, they are inherently ambiguous: many distinct 3D configurations can produce similar images. Depth, on the other hand, offers strong geometric constraints but lacks texture and semantic information. There are also works that train Diffusion Models in 2 stages in various modalities (e.g Optical Flow, Depth, Camera Parameters etc.) \cite{jin2025flovdopticalflowmeets} \cite{Esser_2023_ICCV}\cite{zhong2025flowvlavisualchainthoughtbased} \cite{10.1007/978-3-031-72784-9_4}. A 2 stage pipeline can help to improve controllability as well. However, there does not exist a principled method to train both stages jointly. 

Prior works \cite{alonso2024diffusionworldmodelingvisual} \cite{huang2025vid2world0} \cite{valevski2025diffusion} \cite{hafner2025trainingagentsinsidescalable}\cite{oasis2024} take important steps toward learning diffusion world models from large-scale gameplay data, demonstrating that videos from first-person environments like CS:GO Dust2 and Minecraft offer an ideal testbed for studying long-horizon dynamics. Yet these approaches still rely heavily on action-conditioned modeling or unified representations that must simultaneously encode geometry, appearance, and dynamics, making training fragile, data-hungry, or difficult to scale. A persistent challenge in this space is balancing structural fidelity and visual realism without entangling the two. Methods that treat RGB and depth jointly often struggle to enforce geometric consistency, while methods that train separate models typically require explicit action supervision to maintain temporal coherence. Moreover, action labels are rarely available outside interactive simulators, limiting the applicability of such models to real-world or web-scale video data. 

These limitations raise an important question:
Can we construct a world model that leverages depth to provide stable geometry, while using generative RGB modeling to capture appearance, without requiring action-label supervision throughout? A solution to this would significantly expand the scalability and practicality of diffusion-based world models. It would allow RGB models to be trained on far larger and more diverse datasets, including those without action annotations, while retaining the strong geometric grounding needed for coherent long-horizon rollouts. It would also open the door to training at higher frame rates or using synthetic depth derived from off-the-shelf estimators, thereby bypassing the need for simulator access. In this work, we explore this direction and develop a two-stage diffusion-based world modeling framework that decouples geometry from appearance while maintaining joint predictive capability. Our approach leverages the unique properties of depth and RGB modalities, and is evaluated on the widely used CS:GO Dust2 environment to ensure comparability with prior state-of-the-art systems.

\begin{figure}[H]
    \centering
    \includegraphics[width=0.9\linewidth]{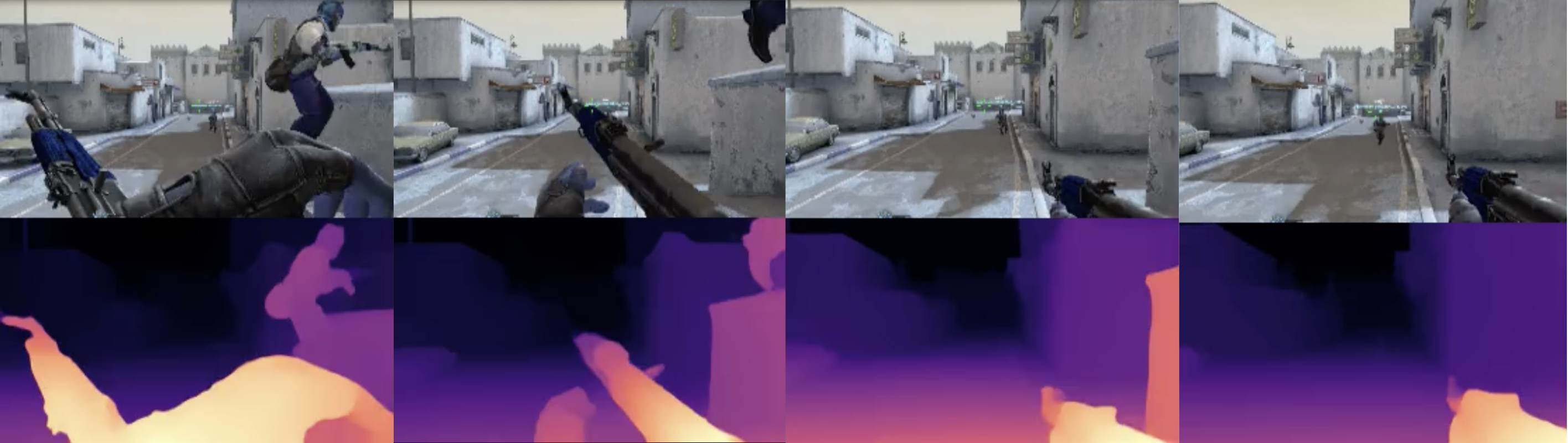}
    \caption{A video sequence of RGB (first row) from the CSGO Dust 2 dataset \cite{10.1109/CoG51982.2022.9893617} and Depth maps (second row) extracted with MiDaS DPT-Hybrid \cite{Ranftl2021}.}
    \label{fig:seq_rgbd}
\end{figure} 
\section{Background}

\paragraph{Diffusion Models.}
Diffusion models \cite{NEURIPS2020_DDPM} have emerged as a powerful class of generative models capable of synthesizing high-fidelity images, videos, and 3D signals. They operate by learning to reverse a fixed Markovian noising process applied to a data sample $\mathbf{z}_0$. The forward diffusion process adds Gaussian noise in $T$ steps:
\begin{equation}
\label{eq:z-param}
\mathbf{z}_t = \sqrt{\bar{\alpha}_t}\,\mathbf{z}_0 
             + \sqrt{1 - \bar{\alpha}_t}\,\boldsymbol{\epsilon}, \boldsymbol{\epsilon} \sim \mathcal{N}(\mathbf{0}, \mathbf{I}), \nonumber
\end{equation}
where $\{\bar{\alpha}_t\}$ denotes the cumulative product of noise schedule coefficients. The generative model learns a denoising function $f_\theta(\mathbf{z}_t, t)$ that estimates either the noise $\boldsymbol\epsilon$, the clean signal $\mathbf{z}_0$, or—following \cite{DBLP:conf/iclr/SalimansH22}—a blended latent variable known as the \emph{velocity} $\mathbf{v}$. The v-prediction parameterization offers improved training stability and sampling efficiency. It reparameterizes Eq \eqref{eq:z-param} in terms of velocity as:
\[
\mathbf{v}_t
= \sqrt{\bar{\alpha}_t}\,\boldsymbol{\epsilon}
  - \sqrt{1-\bar{\alpha}_t}\,\mathbf{z}_0.
\]
This transformation is invertible, and the pair $(\mathbf{z}_t, t)$ uniquely determines $(\mathbf{z}_0, \boldsymbol{\epsilon}, \mathbf{v}_t)$. A neural network then predicts $\hat{\mathbf{v}}_\theta$ in place of $\hat{\boldsymbol\epsilon}_\theta$ or $\hat{\mathbf{z}}_{0,\theta}$ with optional conditioning $\mathbf{c}$ (e.g., actions, class), the denoiser becomes $
\hat{\mathbf{v}}_\theta(\mathbf{z}_t, t \mid \mathbf{c})$ with a simple MSE training objective:
\begin{align}
\mathcal{L}_{\text{v-pred}}
= \mathbb{E}_{\mathbf{z}_0, t, \boldsymbol{\epsilon}}
\Big[
\|
\mathbf{v}_t(\mathbf{z}_0, t)
-
\hat{\mathbf{v}}_\theta(\mathbf{z}_t, t \mid \mathbf{c})
\|^2
\Big].
\end{align}

At sampling time, the predicted velocity is mapped back to the noise or clean-signal estimate. Because the network learns a nearly scale-invariant representation, \textbf{v}-prediction is widely used in high-resolution video diffusion models.

\paragraph{Diffusion forcing for autoregressive prediction.}
To extend diffusion models to sequential or dynamical settings, the model must predict future observations conditioned on past ones. Diffusion forcing \cite{chen2025diffusion, song2025historyguidedvideodiffusion} provides a simple and effective mechanism: instead of predicting a clean future frame $\mathbf{x}_{t+1}$ directly, the model predicts a \emph{noisy} version of it at a randomly sampled diffusion timestep $\tau$. Formally, given a history $\mathcal{H}_t = \{\mathbf{x}_{t-k},\dots,\mathbf{x}_{t}\}$,
\[
(\mathcal{H}_t, \tilde{\mathbf{x}}_{t+1}),
\qquad
\tilde{\mathbf{x}}_{t+1} = q(\mathbf{x}_{t+1},\tau),
\]
is used as a training pair, where $q$ is the forward noising operator. Because the target lies on the diffusion manifold, the model remains close to the support of the forward process even when unrolled autoregressively. This substantially mitigates compounding error and improves long-horizon stability while maintaining parallel training across frames in a given sequence.

\paragraph{Action conditioning in world models.}
World models generalize generative modeling to interactive environments in which future observations depend on both the environment state and agent actions. Diffusion world models incorporate actions by conditioning the denoiser $\hat{\mathbf{v}}_\theta(\mathbf{z}_\tau, \tau \mid \mathcal{H}_t, \mathbf{a}_t),$ which implicitly models $p(\mathbf{x}_{t+1} \mid \mathbf{x}_{1:t}, \mathbf{a}_t).$ Due to the requirement of action-conditioning at each frame, diffusion world models must be temporally autoregressive to see and control the rollout accordingly. Compared to unconditional diffusion models, whose goal is to model a static data distribution, action-conditioned models must learn transition dynamics consistent with control inputs. While effective, this introduces a dependence on action labels, which are often expensive to obtain or unavailable outside simulators. VLAs are also used to predict actions from unlabelled videos but they are often not accurate and require lots of training and supervision. This limitation motivates architectures that either weaken or remove this requirement, enabling training on broader datasets such as augmented videos or web-scale RGB--depth pairs. 
\section{Methodology}
We begin by considering a hypothetical \emph{joint} diffusion transformer (DiT) that models RGB and depth simultaneously. Let
$\mathbf{x}^R_t$ denote the RGB latent at time $t$ and $\mathbf{x}^D_t$ the corresponding depth latent. A naïve joint formulation concatenates the two modalities channel-wise, producing a latent tensor of shape $(C_R+C_D, H, W)$, and learns a denoising function
\[
f_\theta(\mathbf{z}_\tau^{R,D}, \tau \mid \mathbf{z}^{R,D}_{\le t}),
\]
trained with diffusion forcing so that the next RGB latent is conditioned on noisy RGB and depth histories.

However, this joint setting obscures an asymmetry between appearance and geometry: RGB should be \emph{guided} by geometry, whereas depth should be \emph{driven} by dynamics and actions. To expose this structure, we introduce a temporal \textbf{staggering} of modalities. Instead of aligning RGB and depth at the same timestep, we offset depth forward by one frame:
\[
\{\mathbf{x}^R_0, \ldots, \mathbf{x}^R_t\}, \qquad
\{\mathbf{x}^D_1, \ldots, \mathbf{x}^D_{t+1}\}.
\]
This induces two asymmetric conditioning relationships:
\begin{align}
\mathbf{x}^D_{t+1} &\leftarrow (\mathbf{z}^D_{\le t}, \mathbf{z}^R_{\le t-1}, \mathbf{a}_t), \\
\mathbf{x}^R_{t+1} &\leftarrow (\mathbf{z}^R_{\le t}, \mathbf{z}^D_{t+1}). 
\end{align}
Thus the geometric future (depth at $t{+}1$) depends on past geometry, past appearance, and the agent action; but appearance at $t{+}1$ may directly depend on the \emph{current} depth $\mathbf{x}^D_{t+1}$, serving as a geometric guide.

\paragraph{Decoupling into two diffusion stages.}
Instead of training one large joint model, we factor the problem into two specialized DiTs:

\begin{enumerate}
    \item \textbf{Stage 1: Depth DiT (action-conditioned).}\\
    Learns $p(\mathbf{x}^D_{t+1} \mid \mathbf{x}^D_{\le t}, \mathbf{x}^R_{\le t-1}, \mathbf{a}_t)$.  
    This model captures scene geometry and dynamics.
    
    \item \textbf{Stage 2: RGB DiT (depth-guided, action-free).}\\
    Learns $p(\mathbf{x}^R_{t+1} \mid \mathbf{x}^R_{\le t}, \mathbf{x}^D_{t+1})$, enabling RGB synthesis that leverages the depth predicted in Stage~1.
\end{enumerate}

To preserve the training dynamics of a joint model, each DiT receives inputs constructed by concatenating modality channels at matched diffusion noise levels. Specifically, for each training sample we generate noisy latents
\[
\mathbf{z}^R_\tau = q(\mathbf{x}^R, \tau), \qquad
\mathbf{z}^D_\tau = q(\mathbf{x}^D, \tau),
\]
and feed each DiT the same noisy (RGB, depth) pairing it would have seen in the joint model. Diffusion forcing is applied identically in both stages, ensuring that both networks learn to denoise autoregressively on the same manifold of noisy inputs.

\begin{figure}[H]
    \centering
    \includegraphics[width=\linewidth]{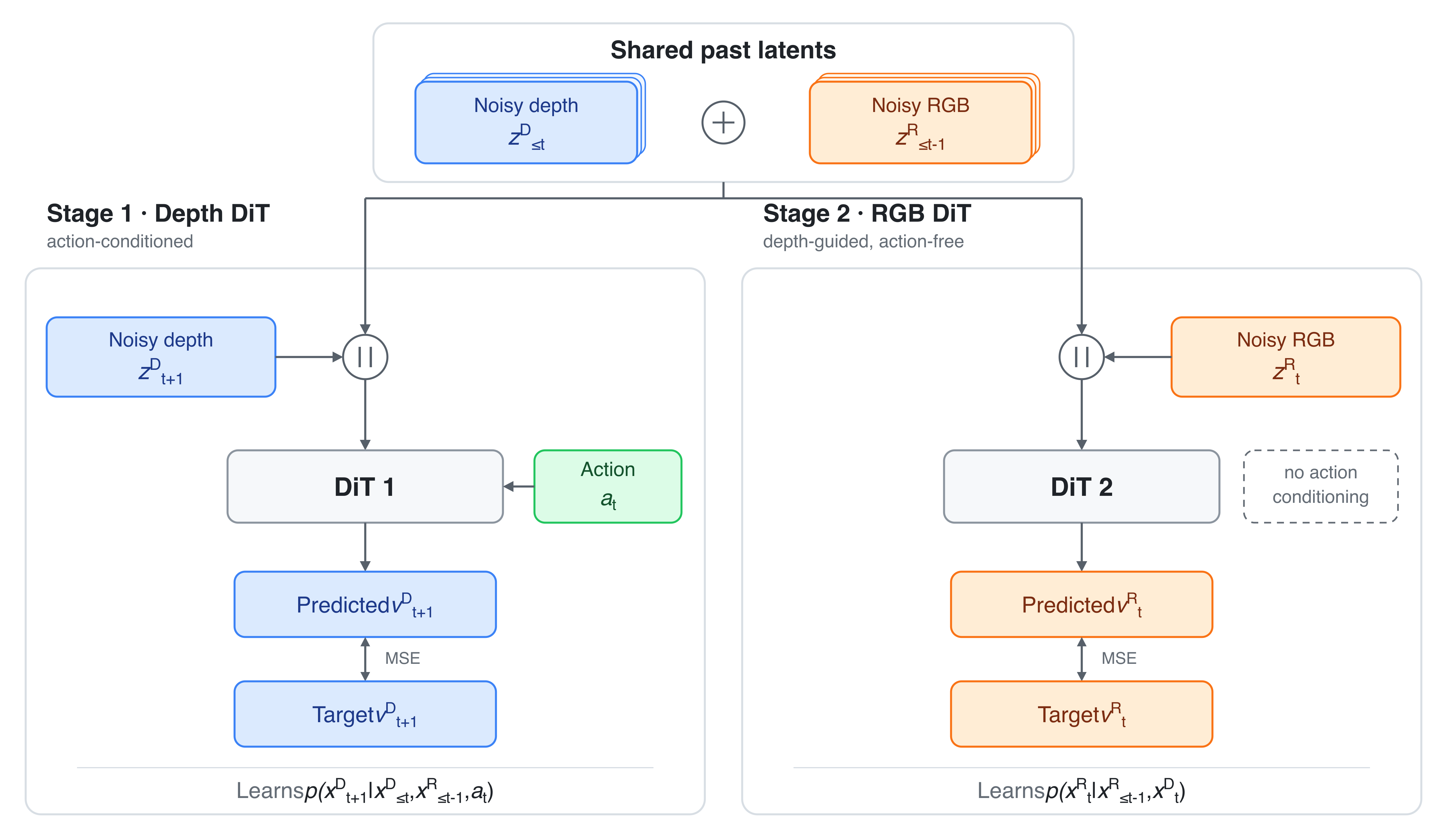}
    \caption{\textbf{Staggered Joint Diffusion Training Strategy.} 
    We illustrate the decoupled training procedure which factorizes the joint modeling of video into two specialized Diffusion Transformers (DiTs). 
    \textbf{Stage 1 (Left):} The Action-Conditioned Depth DiT learns scene dynamics by denoising the depth latent $\mathbf{z}^D_{t+1}$, conditioned on past depth $\mathbf{z}^D_{\le t}$, staggered past RGB $\mathbf{z}^R_{\le t-1}$, and the agent action $\mathbf{a}_t$.
    \textbf{Stage 2 (Right):} The Depth-Guided RGB DiT learns appearance by denoising $\mathbf{z}^R_{t}$. Crucially, it receives structural guidance via the \emph{noisy ground-truth} depth $\mathbf{z}^D_{t}$ (channel-wise concatenated), alongside the full history $\mathbf{z}^{R,D}_{\le t-1}$. 
    Both models utilize v-prediction with a shared noise schedule $\tau$, ensuring the decoupled stages approximate a valid joint distribution.}
    \label{fig:training_pipeline}
\end{figure}

To formally justify our staggered training strategy, we start with the standard autoregressive factorization of the joint distribution for the next frame's RGB ($\mathbf{x}_{t+1}^R$) and depth ($\mathbf{x}_{t+1}^D$). Conditioned on the full history of both modalities and the agent's action $\mathbf{a}_t$, the joint probability decomposes via the chain rule as:
\begin{align}
    p(\mathbf{x}_{t+1}^R, \mathbf{x}_{t+1}^D \mid \mathbf{x}_{1:t}^R, \mathbf{x}_{1:t}^D, \mathbf{a}_t)
    &=
    \underbrace{p(\mathbf{x}_{t+1}^D \mid \mathbf{x}_{1:t}^R, \mathbf{x}_{1:t}^D, \mathbf{a}_t)}_{\text{Dynamics Term}}
    \cdot
    \underbrace{p(\mathbf{x}_{t+1}^R \mid \mathbf{x}_{1:t}^R, \mathbf{x}_{1:t}^D, \mathbf{x}_{t+1}^D, \mathbf{a}_t)}_{\text{Appearance Term}}.
    \label{eq:joint_chain}
\end{align}
Equation~\ref{eq:joint_chain} represents a fully connected joint model. To decouple the training into the specialized stages illustrated in Figure~\ref{fig:training_pipeline}, we introduce two conditional independence assumptions that reflect the causal asymmetry of video generation.

We posit that the physical evolution of the scene (depth at $t{+}1$) is driven by the geometric history and agent actions, rather than the synchronous appearance at time $t$. We therefore assume $\mathbf{x}_{t+1}^D$ is conditionally independent of $\mathbf{x}_{t}^R$:
\begin{equation}
    \mathbf{x}_{t+1}^D \perp \mathbf{x}_{t}^R \mid \mathbf{x}_{1:t-1}^R, \mathbf{x}_{1:t}^D, \mathbf{a}_t.
\end{equation}
This simplifies the dynamics term to $p(\mathbf{x}_{t+1}^D \mid \mathbf{x}_{1:t-1}^R, \mathbf{x}_{1:t}^D, \mathbf{a}_t)$, where the appearance input is staggered by one frame.

We assume that appearance generation is primarily a function of texture history and the target surface geometry. Once the target geometry $\mathbf{x}_{t+1}^D$ is known, the specific action $\mathbf{a}_t$ and past geometry $\mathbf{x}_{1:t}^D$ become redundant. Formally,
\begin{equation}
    \mathbf{x}_{t+1}^R \perp \{\mathbf{x}_{1:t}^D, \mathbf{a}_t\} \mid \mathbf{x}_{1:t}^R, \mathbf{x}_{t+1}^D.
\end{equation}
This simplifies the appearance term to $p(\mathbf{x}_{t+1}^R \mid \mathbf{x}_{1:t}^R, \mathbf{x}_{t+1}^D)$, where the RGB generation is guided by the current depth.

Substituting these simplified terms back into Equation~\ref{eq:joint_chain} yields our final staggered factorization:
\begin{align}
    p(\mathbf{x}_{t+1}^R, \mathbf{x}_{t+1}^D \mid \mathbf{x}_{1:t}^R, \mathbf{x}_{1:t}^D, \mathbf{a}_t)
    =
    \underbrace{p(\mathbf{x}_{t+1}^D \mid \mathbf{x}_{1:t-1}^R, \mathbf{x}_{1:t}^D, \mathbf{a}_t)}_{\text{Stage 1: Staggered Depth DiT}}
    \cdot
    \underbrace{p(\mathbf{x}_{t+1}^R \mid \mathbf{x}_{1:t}^R, \mathbf{x}_{t+1}^D)}_{\text{Stage 2: Depth-Guided RGB DiT}}.
\end{align}

\paragraph{Depth-guided RGB generation.}
During inference, the two stages are executed sequentially. Stage~1 predicts $\mathbf{x}^D_{t+1}$ using action-conditioned rollout. Stage~2 then takes this predicted depth---at the \emph{current} timestep---as an input for RGB generation:
\[
\mathbf{x}^R_{t+1} = f_{\theta_R}(\mathbf{z}^R_\tau, \tau \mid \mathbf{x}^D_{t+1}, \mathbf{x}^R_{\le t}).
\]
Thus RGB can ``peek'' at the depth of the same future frame, yielding a depth-guided appearance model without requiring explicit action-conditioning.

\begin{figure}[H]
    \centering
    \includegraphics[width=0.9\linewidth]{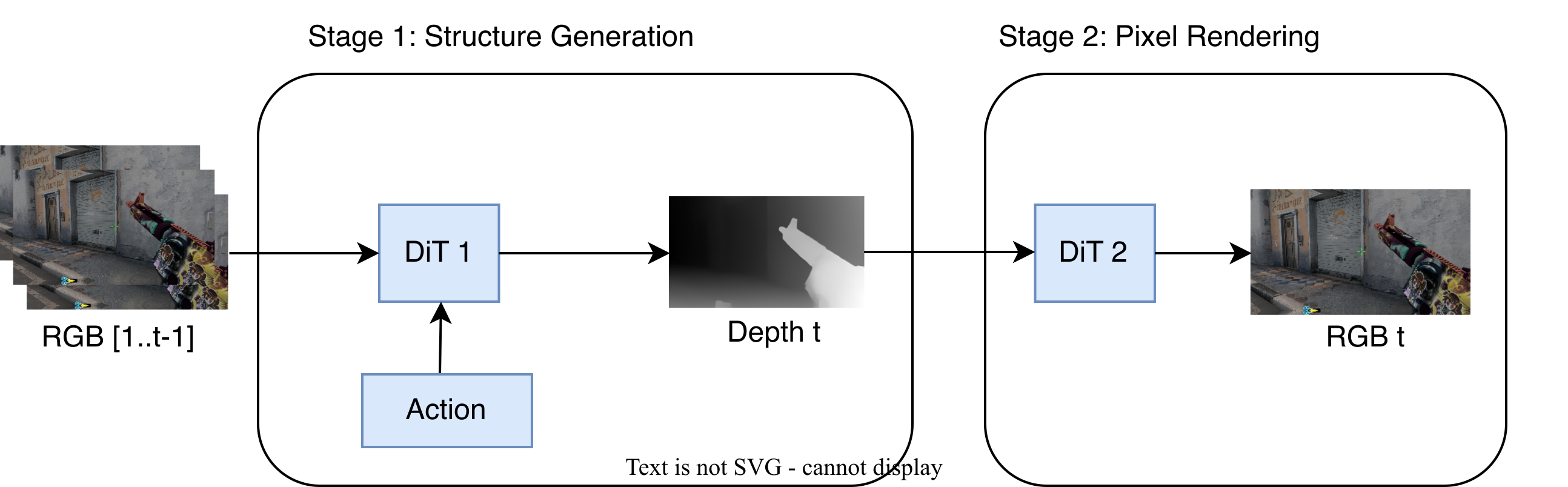}
    \caption{\textbf{Two-Stage Video Synthesis Pipeline}. Stage 1 generates depth maps from the current frame, action and history. Stage 2 renders the next frame conditioned on these history and depth with channel concatenation.}
    \label{fig:pipeline-svg}
\end{figure}

\paragraph{Action-free scalability.}
A key implication of this factorization is that only Stage~1 requires action labels. Stage~2 can be trained or fine-tuned on any dataset with RGB--depth pairs, including augmented data or higher frame rates, with depth obtained from an off-the-shelf estimator. This makes Stage~2 scalable beyond simulator data and enables significantly larger appearance models without requiring interactive trajectories.
\section{Experiments}
\label{sec:experiments}

\subsection{Dataset}

We conduct all experiments on the CS:GO Dust~2 dataset \cite{10.1109/CoG51982.2022.9893617}, which provides large-scale first-person gameplay trajectories suitable for evaluating world models. The dataset consists of approximately 5.5M frames ($\sim$96 hours) recorded at 16~FPS with image resolution $150\times 280$. We apply reflection padding to obtain a uniform resolution of $160\times 280$. We use 5M frames for training and 0.5M frames for validation. Each frame is accompanied by 51-dimensional one-hot encoded actions. As the dataset does not contain ground-truth geometry, we estimate depth using MiDaS DPT-Hybrid \cite{Ranftl2021}.

\subsection{Model Training Setup}

Our models follow the Oasis architecture~\cite{oasis2024} for both the VAE and DiT components, but are retrained to support the Dust~2 resolution and our latent tokenization. All diffusion models operate entirely in the VAE latent space of shape $(16, 28, 16)$.

\subsubsection*{VAE}

The VAE is a ViT-based \cite{MaskedAutoencoders2021} image autoencoder with no temporal compression. Using a patch size of 10, each frame is encoded into a $16\times 28\times 16$ latent tensor. The VAE is trained for 800k iterations with a batch size of 24 (six sequences of four frames) using the following composite objective:
\[
\mathcal{L}
= \alpha \mathcal{L}_{\text{recon}}
+ \beta \mathcal{L}_{\text{LPIPS}}
+ \gamma\, d_{\text{weight}}\mathcal{L}_{\text{GAN}}
+ \lambda \mathcal{L}_{\text{KL}}.
\]
Here $d_{\text{weight}}$ is the dynamic GAN weighting factor,
\[
d_{\text{weight}}
=
\max\!\left(
\frac{\|\nabla \mathcal{L}_{\text{GAN}}\|}
     {\|\nabla \mathcal{L}_{\text{NLL}}\| + 10^{-4}},
\; 10^5
\right),
\]
and we adopt the same balancing scheme as \cite{Rombach_2022_CVPR}. Training is performed in bfloat16 and completes in eight days on a single H100 GPU.

\subsubsection*{Diffusion Transformer (DiT)}

\paragraph{Architecture.}
The diffusion model is a standard DiT~\cite{peebles2023scalablediffusionmodelstransformers} adapted for video generation. Attention is causal along the temporal dimension, and conditioning is injected via AdaLN-Zero. Actions are one-hot encoded and added to the sinusoidal timestep embedding for action-conditioned models. We use noisy ground-truth previous frames as context tokens, following common diffusion-forcing-style conditioning.

\paragraph{Training Procedure.}
Models are trained on sequences of 16 frames with a sigmoid beta noise schedule. Table~\ref{tab:model-config} summarizes the main training configurations.

\paragraph{Inference.}
At inference time, we use DDIM sampling. Given initial context frames $\{\mathbf{x}_t\}_{t=0}^{t_0}$ and a future action sequence $\{\mathbf{a}_t\}_{t=t_0}^{T}$, we encode the context frames into VAE latents, and the DiT autoregressively generates future latents $\{\mathbf{z}_t\}_{t=t_0}^{T}$ consistent with the provided action sequence.

\subsection{Models Evaluated}

To assess the impact of joint RGB--depth modeling and our proposed two-stage formulation, we evaluate five model variants trained under comparable settings. All models use the same VAE tokenizer and DiT backbone described in Section~\ref{sec:experiments}. Unless otherwise stated, training uses action-conditioned diffusion forcing and identical optimization schedules.

\paragraph{RGB-Only Baseline (Action-Conditioned).}
A standard DiT trained purely on RGB latents with one-hot action conditioning. This corresponds to a conventional video diffusion world model without any geometric information.

\paragraph{Joint RGBD Model.}
A single DiT trained on concatenated RGB and depth latents. Both modalities are provided at matching timesteps and are denoised jointly. Actions are used as in the baseline model. This model tests whether depth improves predictive fidelity when fused at the token level.


\paragraph{Two-Stage RGB--Depth Model.}
Our proposed approach factorizes the model into two DiTs trained end-to-end:
\begin{itemize}
    \item \textbf{Stage 1 (Depth DiT).} Predicts future depth conditioned on past depth, past RGB, and actions.
    \item \textbf{Stage 2 (RGB DiT).} Predicts RGB conditioned on past RGB and the \emph{current} predicted depth. No action conditioning is used in Stage~2.
\end{itemize}
We simulate joint-channel diffusion forcing by providing both DiTs with matched noisy RGB and depth latents during training.
We find that our 2-stage pipeline achieves competitive performance with RGB-only, and improved performance over Joint RGBD (as it uses double the number of parameters). With this, however, we are able to have coupled depth and RGB while keeping them decoupled sequentially during generation. We also display samples comparing 2-stage to joint RGBD and Vanilla in Fig.

\begin{table*}[t]
\centering
\caption{Comparison of validation metrics across all model variants. Evaluated on 2400 videos. RGB metrics evaluate appearance fidelity, while depth metrics evaluate geometric accuracy.}
\label{tab:model_comparison}
\begin{tabular}{lccccccccc}
\toprule
\multirow{2}{*}{\textbf{Model}} &
\multicolumn{5}{c}{\textbf{RGB Metrics}} &
\multicolumn{3}{c}{\textbf{Depth Metrics}} \\
\cmidrule(lr){2-6} \cmidrule(lr){7-9}
& FVD$\downarrow$ & FID$\downarrow$ & PSNR$\uparrow$ & LPIPS$\downarrow$ & SSIM$\uparrow$
& AbsRel$\downarrow$ & $\delta_{1.25}\uparrow$ \\
\midrule
Diamond \cite{alonso2024diffusionworldmodelingvisual}                &  368.5 & 87.2 & 18.1 & 0.51 & 0.44 &  -- & -- \\
Vid2World \cite{huang2025vid2world0}               &  106.6 & 17.5 & 18.7 & 0.404 & 0.48 &  -- & -- \\
RGB-Only               &  52.8 & 10.1 & 18.7 & 0.35 & 0.49 &  -- & -- \\
Joint RGBD               &  54.2 & 11.5 & 18.5 & 0.38 & 0.49 &  340.3 & 44.2 \\
2-Stage                  &  53.1 & 11.1 & 18.5 & 0.36 & 0.48 &  318.1 & 44.9 \\
\bottomrule
\end{tabular}
\end{table*}

\subsection{Finetuning RGB DiT on unlabeled data}
We tried take the pretrained two-stage model above, freeze the Stage~1 Depth DiT, and fine-tune only the Stage~2 RGB DiT on additional unlabeled augmented RGB data. We initially settled for augmented data (temporal, flipping, random cropping) because we did not have access to more RGB data. However, we found that this did not improve performance by much. Rather than collecting more RGB data from a game engine, which would have been plausible, we opted instead to simulate our desired setting with a more academic point of view by simply pretraining on a \% of the data and finetuning on the remainder. 
In Fig, we show how reducing the dataset size during training affects the validation metrics similar to the findings from GameNGen. In Fig, we show how finetuning only the RGB DiT on unlabeled RGB data can improve the quality despite the Depth DiT only trained on some \% of the data. This corroborates our hypothesis (also CITE) that learning depth is more sample efficient for the model. Additionally, the training loss is also an order of magnitude lower.

\section{Limitations and Future Work}
Some of the limitations of our work that need to be addressed are:
\begin{itemize}
    \item More ablations needed while varying model sizes.
    \item While we need less action labels, we require estimating (imperfect) depth maps or acess to ground truth depth maps which can be a big ask. Imperfect depth learnt renders the Depth DiT useless.
    \item We have 2 separate models and that might result in more accumulative error. There is also a distribution mismatch between the two DiTs due to the frame offset.
\end{itemize}
Some extensions of our work may include:
\begin{itemize}
    \item Incorporating other modalities such as camera parameters, occlusion maps, and optical flow maps.
    \item Being able to generalize to arbitrary OOD datasets which would require a much larger scale model with a more diverse dataset.
    \item Scaling depth prediction using multi-view or self-supervised priors, extending our approach to real-world video with estimated camera poses, and exploring hierarchical or multi-rate variants of two-stage diffusion for long-horizon planning and embodied agents.    
\end{itemize}

\section{Conclusion}
We present a two-stage diffusion world model that decouples geometric and appearance prediction through a staggered RGB–depth formulation. By training a depth-predictive DiT with action conditioning and an RGB-predictive DiT that is guided by the current depth but free of action inputs, we showed that the factorized model retains the expressiveness of a joint RGBD diffusion model while enabling practical advantages in scalability and data efficiency. Our formulation is theoretically grounded through its correspondence to the conditionals learned by a fully joint diffusion model under diffusion forcing, and empirically validated on the CS:GO Dust~2 dataset, where it improves RGB fidelity, depth accuracy, and long-horizon rollout stability relative to traditional baselines. A key benefit of our approach is that the RGB model can be fine-tuned on arbitrary RGB–depth data, including augmented or unlabeled sources, since it requires no action supervision. This opens the door to substantially larger training corpora and richer generative models of appearance, while maintaining geometric consistency through the depth stage.
\section*{Acknowledgements}
We thank Esmeralda Whitammer for her helpful discussions on VAE training. Cameron Barker for his useful suggestions on attention masking. Danier Duolikun for his helpful advice on video diffusion modeling papers and guidance. This work was done during an internship at Qubit Analytics.
\bibliographystyle{plain}
\bibliography{References/references}

@misc{jin2025flovdopticalflowmeets,
      title={FloVD: Optical Flow Meets Video Diffusion Model for Enhanced Camera-Controlled Video Synthesis}, 
      author={Wonjoon Jin and Qi Dai and Chong Luo and Seung-Hwan Baek and Sunghyun Cho},
      year={2025},
      eprint={2502.08244},
      archivePrefix={arXiv},
      primaryClass={cs.CV},
      url={https://arxiv.org/abs/2502.08244}, 
}

@InProceedings{Esser_2023_ICCV,
    author    = {Esser, Patrick and Chiu, Johnathan and Atighehchian, Parmida and Granskog, Jonathan and Germanidis, Anastasis},
    title     = {Structure and Content-Guided Video Synthesis with Diffusion Models},
    booktitle = {Proceedings of the IEEE/CVF International Conference on Computer Vision (ICCV)},
    month     = {October},
    year      = {2023},
    pages     = {7346-7356}
}

@misc{zahorodnii2025improvingworldmodelsusing,
      title={Improving World Models using Deep Supervision with Linear Probes}, 
      author={Andrii Zahorodnii},
      year={2025},
      eprint={2504.03861},
      archivePrefix={arXiv},
      primaryClass={cs.AI},
      url={https://arxiv.org/abs/2504.03861}, 
}

@article{oasis2024,
  author    = {Decart and Quevedo, Julian and McIntyre, Quinn and Campbell, Spruce and Chen, Xinlei and Wachen, Robert},
  title     = {Oasis: A Universe in a Transformer},
  year      = {2024},
  url       = {https://oasis-model.github.io/}
}

@misc{peebles2023scalablediffusionmodelstransformers,
      title={Scalable Diffusion Models with Transformers}, 
      author={William Peebles and Saining Xie},
      year={2023},
      eprint={2212.09748},
      archivePrefix={arXiv},
      primaryClass={cs.CV},
      url={https://arxiv.org/abs/2212.09748}, 
}

@InProceedings{aether,
    author    = {Zhu, Haoyi and Wang, Yifan and Zhou, Jianjun and Chang, Wenzheng and Zhou, Yang and Li, Zizun and Chen, Junyi and Shen, Chunhua and Pang, Jiangmiao and He, Tong},
    title     = {Aether: Geometric-Aware Unified World Modeling},
    booktitle = {Proceedings of the IEEE/CVF International Conference on Computer Vision (ICCV)},
    month     = {October},
    year      = {2025},
    pages     = {8535-8546}
}

@InProceedings{tesseract,
    author    = {Zhen, Haoyu and Sun, Qiao and Zhang, Hongxin and Li, Junyan and Zhou, Siyuan and Du, Yilun and Gan, Chuang},
    title     = {Learning 4D Embodied World Models},
    booktitle = {Proceedings of the IEEE/CVF International Conference on Computer Vision (ICCV)},
    month     = {October},
    year      = {2025},
    pages     = {5337-5347}
}

@inproceedings{alonso2024diffusionworldmodelingvisual,
      title={Diffusion for World Modeling: Visual Details Matter in Atari},
      author={Eloi Alonso and Adam Jelley and Vincent Micheli and Anssi Kanervisto and Amos Storkey and Tim Pearce and François Fleuret},
      booktitle={Thirty-eighth Conference on Neural Information Processing Systems}}

@article{huang2025vid2world0,
      title   = {Vid2World: Crafting Video Diffusion Models to Interactive World Models},
      author  = {Siqiao Huang and Jialong Wu and Qixing Zhou and Shangchen Miao and Mingsheng Long},
      year    = {2025},
      journal = {arXiv preprint arXiv: 2505.14357}
    }

@inproceedings{
valevski2025diffusion,
title={Diffusion Models Are Real-Time Game Engines},
author={Dani Valevski and Yaniv Leviathan and Moab Arar and Shlomi Fruchter},
booktitle={The Thirteenth International Conference on Learning Representations},
year={2025},
url={https://openreview.net/forum?id=P8pqeEkn1H}
}

@misc{hafner2025trainingagentsinsidescalable,
      title={Training Agents Inside of Scalable World Models}, 
      author={Danijar Hafner and Wilson Yan and Timothy Lillicrap},
      year={2025},
      eprint={2509.24527},
      archivePrefix={arXiv},
      primaryClass={cs.AI},
      url={https://arxiv.org/abs/2509.24527}, 
}

@article{yan2025,
  title   = {Yan: Foundational Interactive Video Generation},
  author  = {Ye, Deheng and Zhou, Fangyun and Lv, Jiacheng and Ma, Jianqi and Zhang, Jun and Lv, Junyan and Li, Junyou and Deng, Minwen and Yang, Mingyu and Fu, Qiang and Yang, Wei and Lv, Wenkai and Yu, Yangbin and Wang, Yewen and Guan, Yonghang and Hu, Zhihao and Fang, Zhongbin and Sun, Zhongqian},
  journal = {arXiv preprint arXiv:2508.08601},
  year    = {2025}
}

@misc{chen2025deepverse4dautoregressivevideo,
      title={DeepVerse: 4D Autoregressive Video Generation as a World Model}, 
      author={Junyi Chen and Haoyi Zhu and Xianglong He and Yifan Wang and Jianjun Zhou and Wenzheng Chang and Yang Zhou and Zizun Li and Zhoujie Fu and Jiangmiao Pang and Tong He},
      year={2025},
      eprint={2506.01103},
      archivePrefix={arXiv},
      primaryClass={cs.CV},
      url={https://arxiv.org/abs/2506.01103}, 
}

@misc{zhong2025flowvlavisualchainthoughtbased,
      title={FlowVLA: Visual Chain of Thought-based Motion Reasoning for Vision-Language-Action Models}, 
      author={Zhide Zhong and Haodong Yan and Junfeng Li and Xiangchen Liu and Xin Gong and Tianran Zhang and Wenxuan Song and Jiayi Chen and Xinhu Zheng and Hesheng Wang and Haoang Li},
      year={2025},
      eprint={2508.18269},
      archivePrefix={arXiv},
      primaryClass={cs.RO},
      url={https://arxiv.org/abs/2508.18269}, 
}

@inproceedings{10.1007/978-3-031-72784-9_4, author = {Liang, Jingyun and Fan, Yuchen and Zhang, Kai and Timofte, Radu and Van Gool, Luc and Ranjan, Rakesh}, title = {MoVideo: Motion-Aware Video Generation with Diffusion Model}, year = {2024}, isbn = {978-3-031-72783-2}, publisher = {Springer-Verlag}, address = {Berlin, Heidelberg}, url = {https://doi.org/10.1007/978-3-031-72784-9_4}, doi = {10.1007/978-3-031-72784-9_4}, booktitle = {Computer Vision – ECCV 2024: 18th European Conference, Milan, Italy, September 29–October 4, 2024, Proceedings, Part XLIV}, pages = {56–74}, numpages = {19}, location = {Milan, Italy} }

@inproceedings{10.1109/CoG51982.2022.9893617,
author = {Pearce, Tim and Zhu, Jun},
title = {Counter-Strike Deathmatch with Large-Scale Behavioural Cloning},
year = {2022},
publisher = {IEEE Press},
url = {https://doi.org/10.1109/CoG51982.2022.9893617},
doi = {10.1109/CoG51982.2022.9893617},
booktitle = {2022 IEEE Conference on Games (CoG)},
pages = {104–111},
numpages = {8},
location = {Beijing, China}
}

@article{Ranftl2021,
	author    = {Ren\'{e} Ranftl and Alexey Bochkovskiy and Vladlen Koltun},
	title     = {Vision Transformers for Dense Prediction},
	journal   = {ICCV},
	year      = {2021},
}

@article{chen2025diffusion,
  title={Diffusion forcing: Next-token prediction meets full-sequence diffusion},
  author={Chen, Boyuan and Mart{\'\i} Mons{\'o}, Diego and Du, Yilun and Simchowitz, Max and Tedrake, Russ and Sitzmann, Vincent},
  journal={Advances in Neural Information Processing Systems},
  volume={37},
  pages={24081--24125},
  year={2025}
}

@misc{song2025historyguidedvideodiffusion,
  title={History-Guided Video Diffusion}, 
  author={Kiwhan Song and Boyuan Chen and Max Simchowitz and Yilun Du and Russ Tedrake and Vincent Sitzmann},
  year={2025},
  eprint={2502.06764},
  archivePrefix={arXiv},
  primaryClass={cs.LG},
  url={https://arxiv.org/abs/2502.06764}, 
}

@inproceedings{DBLP:conf/iclr/SalimansH22,
  author       = {Tim Salimans and
                  Jonathan Ho},
  title        = {Progressive Distillation for Fast Sampling of Diffusion Models},
  booktitle    = {The Tenth International Conference on Learning Representations, {ICLR}
                  2022, Virtual Event, April 25-29, 2022},
  publisher    = {OpenReview.net},
  year         = {2022},
  url          = {https://openreview.net/forum?id=TIdIXIpzhoI},
  bibsource    = {dblp computer science bibliography, https://dblp.org}
}

@inproceedings{NEURIPS2020_DDPM,
  author = {Ho, Jonathan and Jain, Ajay and Abbeel, Pieter},
  booktitle = {Advances in Neural Information Processing Systems},
  editor = {Larochelle, H. and Ranzato, M. and Hadsell, R. and Balcan, M.F. and Lin, H.},
  pages = {6840--6851},
  publisher = {Curran Associates, Inc.},
  title = {Denoising Diffusion Probabilistic Models},
  volume = 33,
  year = 2020
}

@Article{MaskedAutoencoders2021,
  author  = {Kaiming He and Xinlei Chen and Saining Xie and Yanghao Li and Piotr Doll{\'a}r and Ross Girshick},
  journal = {arXiv:2111.06377},
  title   = {Masked Autoencoders Are Scalable Vision Learners},
  year    = {2021},
}

@InProceedings{Rombach_2022_CVPR,
    author    = {Rombach, Robin and Blattmann, Andreas and Lorenz, Dominik and Esser, Patrick and Ommer, Bj\"orn},
    title     = {High-Resolution Image Synthesis With Latent Diffusion Models},
    booktitle = {Proceedings of the IEEE/CVF Conference on Computer Vision and Pattern Recognition (CVPR)},
    month     = {June},
    year      = {2022},
    pages     = {10684-10695}
}
\appendix
\section{Model Configuration}
\begin{table}[H]
\centering
\caption{Model configuration for both DiTs.}
\label{tab:model-config}
\begin{tabular}{ll}
\toprule
\textbf{Component} & \textbf{Specification} \\
\midrule
Number of parameters   & 600M \\
Batch size             & 24 \\
Sequence length        & 16 \\
EMA decay              & 0.999 \\
Optimizer              & Adafactor (lr = $1\times 10^{-4}$) \\
Precision              & bfloat16 \\
Architecture           & DiT \\
Patch size             & 2 \\
GPU                    & NVIDIA GH200 \\
Training time          & $\sim$2 days \\
\bottomrule
\end{tabular}
\end{table}

\end{document}